\documentclass[conference, a4paper]{IEEEtran}

\usepackage{float}

\usepackage{graphics}
\usepackage{graphicx}
\usepackage{subcaption}

\ifCLASSOPTIONcompsoc
  \usepackage[nocompress]{cite}
\else
  \usepackage{cite}
\fi

\ifCLASSINFOpdf
\else
\fi
\newcommand{\code}{\texttt}
\usepackage{xcolor}
\usepackage{comment}
\usepackage{booktabs}
\usepackage{amsmath}
\usepackage{amssymb}
\usepackage{stmaryrd}
\usepackage{pifont}
\usepackage{color, colortbl}

\newcommand{\cmark}{\ding{51}}
\newcommand{\xmark}{\ding{55}}

\newcommand{\fakepar}[1]{~\\\noindent\textbf{#1.}}
\newcommand{\firstfakepar}[1]{\noindent\textbf{#1.}}

\begin{document}


\title{Rare Yet Popular: Evidence and Implications from\\Labeled Datasets for Network Anomaly Detection}

\author{
\IEEEauthorblockN{Jose Manuel Navarro, Alexis Huet, Dario Rossi}
\IEEEauthorblockA{Huawei Technologies, co. Ltd}
\texttt{\{jose.navarro,alexis.huet,dario.rossi\}@huawei.com}
}

\maketitle
\pagestyle{plain} 
\bstctlcite{IEEEexample:BSTcontrol}
\begin{abstract}

Anomaly detection research works generally propose algorithms or end-to-end systems that are designed to automatically discover outliers in a dataset or a stream. While literature abounds concerning algorithms or the definition of metrics for better evaluation, the quality of the ground truth against which they are evaluated is seldom questioned. In this paper, we present a systematic analysis of available public (and additionally our private) ground truth for anomaly detection in the context of network environments, where data is intrinsically temporal, multivariate and, in particular, exhibits \emph{spatial} properties, which, to the best of our knowledge, we are the first to explore. Our analysis reveals that, while anomalies are, by definition, temporally rare events, their spatial characterization clearly shows \emph{some type of anomalies are significantly more popular than others}. 
We find that simple clustering can reduce the need for human labeling by a factor of 2$\times$-10$\times$, that we are first to quantitatively analyze in the wild.


\end{abstract}

\IEEEpeerreviewmaketitle

\section{Introduction}\label{sec:intro}
Troubleshooting network anomalies is a fundamental task of modern network management.  The overall workflow may require to analyze categorical (e.g., configuration information), textual (e.g., system or service logs),  numeric (e.g., Key Performance Indicators, KPI, telemetry) information, or a multi-modal combination of the former, to identify and repair the fault. 
Yet, current networks have achieved a size and complexity that renders human-based Anomaly Detection (AD) and diagnosis unfeasible without any kind of automated help~\cite{bhuyan2013network}:  in this regard, Machine Learning (ML) has been consistently exploited~\cite{ahmed2016survey} to efficiently solve the AD task.

In particular, we observe that the emergence of model driven telemetry supported by multiple vendors  has made accessible a wealth of (i) \emph{time-varying} and (ii) \emph{multivariate} (iii) \emph{numerical} data: consequently, 
literature to perform AD in these settings has flourished in recent times. First, an abundant \emph{corpus of ML algorithms} and methods has been investigated -- from changepoint detection~\cite{truong2020selective}, to
trees~\cite{if08icdm,hariri2019extended}, subspace projection~\cite{pevny2016loda,manzoor2018xstream} deep learning~\cite{kwon2019survey},   to name a few-- though not  surprisingly, there is no single method that appears to be superior to all competitors in all circumstances.  Additionally, as many of the above approaches do not explicitly take into account the temporal component of multivariate KPIs data,  a complementary research track~\cite{singh2017demystifying,xu2018unsupervised,tatbul2018precision, wu2021current} has started exploring \emph{unbiased metrics for AD evaluation}  on temporal data, to avoid  potentially misleading or downright incorrect results.
Under this light, regardless of the ML algorithm class, proper evaluation of AD algorithms still requires some amount of \emph{labeled data} -- which holds also for  unsupervised algorithms, where labels are used not for the purpose of model training, but rather as \emph{ground truth} (GT) for a fair algorithmic assessment.
As, by definition, anomalous events are rare and as data labeling is, additionally, a knowingly difficult task, a third research track is devoted to assessing the label quality~\cite{wu2021current} or reducing the burden of ground truth collection via e.g., active-learning~\cite{active-learning,zhao2019label}  or synthetic generation~\cite{tsagen21tnsm}.

At the same time, whereas as early highlighted network troubleshooting is inherently multivariate, recent work on metrics~\cite{tatbul2018precision, wu2021current} or ground truth~\cite{active-learning,zhao2019label,tsagen21tnsm} evaluation is still focused on \emph{univariate} data -- which is partly due to the lack, up to  recent times, of publicly available  datasets.
To counter this problem, we set out to systematically analyze and contrast the available datasets for time-varying multivariate anomaly detection, with special attention to the computer network domain. In particular, instead of proposing yet another ``new'' algorithm to an already 
quite extensive arsenal (e.g., see surveys in~\cite{bhuyan2013network,ahmed2016survey,kwon2019survey}), 
we focus on the study of the \emph{ground truth}   labels, of which we analyze its \emph{temporal} (i.e., which period is labeled as anomalous) and \emph{spatial}  (i.e., which KPIs are labeled as anomalous during that interval) properties. We point out that, due to the lack of public data until recent times, \emph{spatial GT properties have not been studied yet}.

Summarizing our main contributions, we first present a thorough analysis  of GT under  temporal  (e.g., outlier vs  long lasting anomaly) and spatial (univariate vs multivariate anomaly) angles. The presence of (i) different types of temporal anomalies 
with a, furthermore, (ii)  intrinsically multivariate GT labels suggests the use of complementary detection techniques (e.g., outlier vs changepoint detection), and stresses the need of local outlier explanation in the spatial dimension. 
Second, we leverage binary clustering to group spatially similar anomalies and automatically  provide human-readable cluster labels in the original KPI space, that can greatly assist the labeling tasks (e.g., in active learning settings). Third, we analyze the clustering results, empirically quantifying that, despite anomalies being temporally rare, the spatial footprint of some is significantly more popular than others -- with e.g., top-5 (10) clusters representing the bulk 75\% (85\%) of anomalous events in the analyzed datasets,   \emph{based on which we  quantify  expected benefits of cluster-based active learning in the wild}. 
\begin{table*}[t]
\caption{Statistical description of the data sources used in this paper.}\label{tab:datasets}
\centering
\begin{tabular}{lcccccccc}
\toprule
\begin{tabular}[c]{@{}l@{}}\textbf{Data}\\ \textbf{Source}\end{tabular} &
\textbf{Public} & 
\begin{tabular}[c]{@{}c@{}}\textbf{No. of}\\ \textbf{datasets}\end{tabular} &
\begin{tabular}[c]{@{}c@{}}\textbf{Sampling}\\ \textbf{period}\end{tabular} & \textbf{\begin{tabular}[c]{@{}c@{}}\textbf{Median$\pm$MAD$^\dagger$ }\\ \textbf{samples / dataset}\end{tabular}} & \begin{tabular}[c]{@{}c@{}}\textbf{Median$\pm$MAD$^\dagger$ }\\ \textbf{features / dataset}\end{tabular} & \begin{tabular}[c]{@{}c@{}}\textbf{No. of}\\ \textbf{events}\end{tabular} & \begin{tabular}[c]{@{}l@{}}\textbf{Median$\pm$MAD$^\dagger$}\\ \textbf{\# KPIs / event}\end{tabular} &
\begin{tabular}[c]{@{}l@{}}\textbf{Ground truth$^\ddagger$}\\ \textbf{Type, Extent}\end{tabular} 
\\ 
\midrule
\multicolumn{1}{l}{BGP~\cite{dataset_cisco}}        & \cmark          & 39                   & 5 sec                                                                    & \hspace{4pt}1K $\pm$ 31                                                                                              & \hspace{1pt} 873 $\pm$ 18                                                                                     & 45                                                                          & n.a.                                                                        & controlled, temporal \\
\multicolumn{1}{l}{Webserver~\cite{dataset_kaggle}} & \cmark          & 15                   & 1 min                                                                     & 398K $\pm$ 146                                                                                           & 231 $\pm$ 0                                                                                           & 64                                                                          & n.a.                                                                        & \hspace{5pt}manual, temporal   \\
\midrule
\multicolumn{1}{l}{Router}                   & \xmark          & 64                   & 1 min                                                                & \hspace{4pt}2K $\pm$ 1K                                                                                              & \hspace{4pt} 87 $\pm$ 79                                                                                      & 175                                                                         & 4 $\pm$ 4.4                                                               & \hspace{5pt}manual, spatio-temporal     \\
\multicolumn{1}{l}{SMD~\cite{dataset_omni}}         & \cmark          & 28                   & 1 min                                                                   & 47K $\pm$ 37                                                                                             & \hspace{4pt}38  $\pm$ 0                                                                                             & 325                                                                         & 4 $\pm$ 2.9                                                              & \hspace{5pt}manual, spatio-temporal  \\
\bottomrule
\multicolumn{9}{l}{$\dagger$ MAD stands for Median Absolute Deviation. $\ddagger$  GT Type is either controlled or manual, GT extent is either temporal or spatio-temporal.}\\


\end{tabular}
\end{table*}
\begin{figure*}[t]
    \centering
    \includegraphics[width=\textwidth]{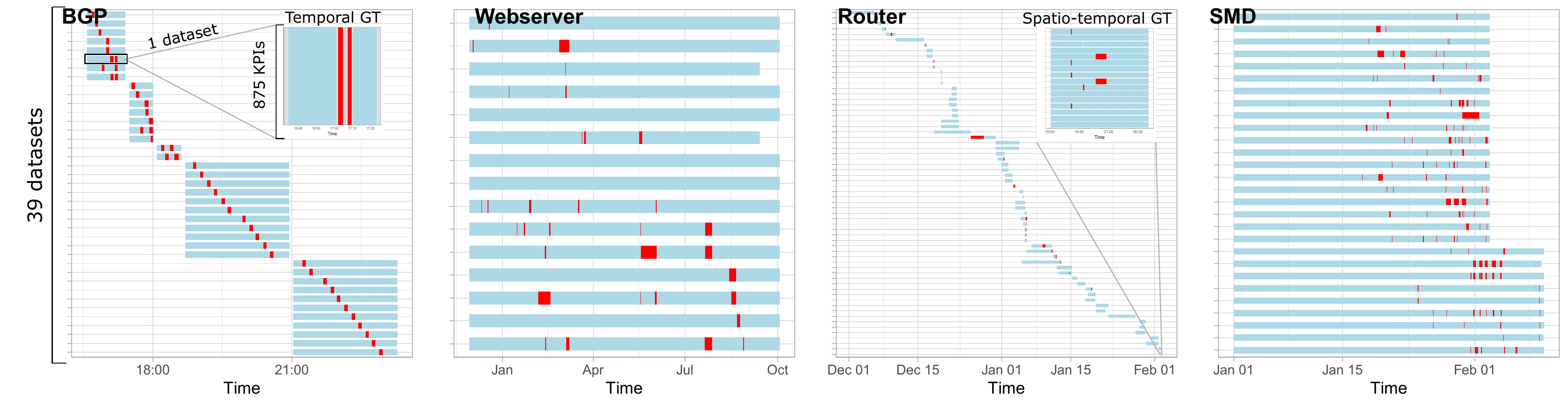}
    \caption{\emph{Data sources}: Temporal representation, where each blue line represents a dataset, with labeled anomalies marked in red.  The inset plots of BGP and Router further expand a single dataset (time on x-axis, KPI on the y-axis) to portray differences of temporal GT (BGP, no per-KPI information) and complete GT (Router, detailed per-timeslot per-KPI information). 
    }
    \label{fig:temporal_evolution}
\end{figure*}

The remainder of this paper describes the data (Sec.~\ref{sec:dataset}), analyzes temporal  (Sec.~\ref{sec:result_time})  and spatial (Sec.~\ref{sec:result_space})  properties of the GT and summarizes our findings (Sec.~\ref{sec:discuss}).

\section{Data}\label{sec:dataset}

We use four different sources of data, which to the best of our knowledge comprise all \emph{publicly} available data with labeled ground truth related to  (i) time-varying (ii) multivariate (iii) numerical data in the computer networks\footnote{Whereas other publicly available sources exist (e.g., water treatment~\cite{goh2016dataset, shin2020hai}), we disregard them in this paper for the sake of simplicity.} domain. We additionally consider a \emph{private} data source, that we cannot, unfortunately, share due to legal aspects. 

\fakepar{Overall summary}
As summarized in Tab.~\ref{tab:datasets}, these data sources have highly heterogeneous characteristics. Each source contains several  \emph{datasets}, each of which is  a single instance of multivariate numeric time series (e.g., collected on a specific device in the topology),
that we  visually depict  at a glance  in  Fig.~\ref{fig:temporal_evolution}, where each blue line represents a different dataset\footnote{Temporal gaps between datasets in the BGP and Router sources have been reduced for visual clarity, as the pictures would have been too sparse otherwise.} and labeled anomalies are marked in red.

From a \emph{temporal} perspective, datasets are sampled at minute (all but BGP) or second (BGP) granularity, with duration ranging from 
thousands (BGP, Router) to hundreds of thousands (Webserver) of samples. For each source, datasets can be either systematically collected at the same (i.e., Webserver) or only partly overlapping (i.e., BGP, Router) or unknown (i.e., SMD) times.  From a \emph{spatial} perspective, the set of collected  KPIs (or features)  ranges from few tens (Router, SMD) to few hundreds (BGP, Webserver). Additionally, within each dataset, the spatial dimension is either fixed (Webserver, SMD) or variable (BGP, Router). 

From a  \emph{GT labeling} perspective, the number of anomalous events ranges between few tens (BGP, Webserver) to few hundreds (Router, SMD). Most importantly, we can categorize the available ground truth along the (i)  type  and (ii)  extent.
In terms of type, anomalies are either \emph{controlled} by a synthetic injection process (BGP) or \emph{manually} verified by human experts (all but BGP). In terms of extent, the label can only provide a \emph{temporal} view (BGP, Webserver) of the anomalous event, or  a complete \emph{spatio-temporal} view (Router, SMD) by precisely pinpointing which KPIs exhibit an anomalous behavior during the interval. As a result, for some data sources (Router, SMD) it is possible to quantify the spatial footprint of an anomaly (e.g., median KPI/event as in Tab.~\ref{tab:datasets}) and further analyze its prevalence (e.g., clustering as in Sec.~\ref{sec:result_space}).  
We now provide a brief overview for each source.

\fakepar{BGP} A public~\cite{dataset_cisco} collection of 39 datasets, this source contains multiple experiments run on a testbed datacenter with 8 leaf nodes connected by 4 spine nodes: a dataset in this source represents a specific device of the deployed testbed. 
The testbed runs only BGP as a routing protocol in the control plane, with application traffic in the data plane peaking up to 1\,Tbps. The collected KPIs are selected by experts, and relate to either control plane (i.e., BGP protocol telemetry) or data plane (e.g., per node and per interface counters)  features. 

Ground truth is \emph{controlled} and \emph{temporal}, with two types of injected anomalies (BGP clear; BGP port flap).
GT labels reliably indicate the start time of the injected BGP anomaly, however the anomaly duration is arbitrarily set to a duration of 5 minutes (equivalent to 60 samples due to the 5 seconds sampling period). Additionally, the ground truth reliably indicates  the device (equivalently, dataset) where the anomaly has been injected, but no manual annotation is further performed. Yet, from  Fig.~\ref{fig:temporal_evolution},  five different experimental groups are clearly visible, with a controlled pattern of sequential anomaly injections across the different devices. Thus, when working with datasets from this source, one should expect \emph{unlabeled} anomalies to be also present -- due to cascading effects from neighboring devices.

\fakepar{Webserver} A public~\cite{dataset_kaggle} collection of 15 datasets, collected from multiple servers hosting an enterprise application. KPIs belonging to several categories (e.g., database connections, memory, transactions, threads, swap and other) are collected from the operating system and the WebLogic Server application.  Individual datasets in this source represent distinct hosts, that may still exhibit correlated behaviors due to the enterprise application dependencies: e.g.,  several moments in time exhibiting simultaneous anomalies are visible in Fig.~\ref{fig:temporal_evolution}.

The ground truth is \emph{temporal} and \emph{manual}, i.e., experts analyzed the moments in which the application was restarted and labeled the time period before the restart as either anomalous or healthy. Since experts validate only a subset of the temporal period (i.e., the time before a restart), it follows that every other period that does not precede an application restart is labeled as normal -- which may mask statistically anomalous data with less prominent business impact.

\fakepar{Router}  A private  collection of 64 datasets, each of which  correspond to a single router of a Internet Service Provider (ISP)  deployment. Generally, only up to two
datasets are  extracted from the same deployment and exhibit temporal overlap,  while most of the datasets spans different ISPs and are thus independent.  Each dataset presents a variable number of KPIs, whose name comprises two parts: a detailed node/module/chassis identifier (which we disregard) and a KPI identifier (for spatial analysis).  

The  ground truth collection is \emph{manual} and \emph{spatio-temporal}. As in Webserver, an alert ticket prompts human experts to examine a set of KPIs, to identify anomalous vs healthy periods. In addition to Webserver, the experts additionally flag each KPI individually as being  anomalous or not: the number of KPI in each dataset depends on the set of KPIs examined by each expert and is significantly more variable with respect to other sources. To make the spatio-temporal analysis more general in Sec.~\ref{sec:result_space}, we consider each name as ``flat'', and do not attempt at using semantic information (e.g., such as KPI category) or syntactical structure (e.g., hierarchical naming structure).

\fakepar{SMD}  Server Machine Dataset (SMD) is a public~\cite{dataset_omni} collection of 28 datasets, each of which contains performance metrics like  CPU, memory or network usage. While the servers were deployed in three groups, each dataset must be treated independently from each other. We additionally remark that whereas Fig.~\ref{fig:temporal_evolution} visually depicts them as perfectly simultaneous, this does not correspond to the real moments in time in which each dataset was collected, as this information is not present in the available data. 

The ground truth for these datasets is, as in the previous case,
\emph{manual} and \emph{spatio-temporal}, with human experts examining  incident reports and  flagging  KPIs individually. 
Roughly the first half of the data does not exhibit any anomalous KPI (which is convenient for learning self-supervised representations of the data~\cite{dataset_omni}), in stark contrast with the second half of the data (that exhibits a significantly large number of anomalies).
Unlike in the Router source, no identifier information is present in KPI names and, coherently with the Router case, we consider each name as ``flat'' to preserve generality of the spatial analysis in  Sec.~\ref{sec:result_space}. 

\fakepar{Insights for anomaly detection}
The brief description in this section already highlights important facts, concerning the released ground truth, that have to be taken into account to perform anomaly detection in a way that produces results that are correct and true to the underlying data. 

First, incompleteness in the GT or meta-data can lead to biased evaluation. For instance, cascading effect among neighboring nodes may arise that are not properly documented (e.g., SMD source lacks topological meta-data) and labeled (e.g., BGP source only labels the injection node, but do not label affected nodes). Similarly, bias can be introduced by temporal \emph{incompleteness}  (e.g., as experts may purposely avoid to label anomalies that are irrelevant from business perspective) or  \emph{approximation} (e.g., when anomaly duration is fixed in controlled labeling). Finally, GT may exhibit spatial sparseness due to labeling incompleteness (e.g., false negative of an expert missing a statistically anomalous KPI), or different tacit purposes of GT labeling (e.g., labeling the cause vs labeling all symptoms).    

In the remainder of this work, we more systematically analyze GT from a spatio-temporal angle, to gather richer insights and implications concerning the network AD task.

\section{Temporal Properties}\label{sec:result_time}
\begin{figure*}[h]
    \centering
    \includegraphics[width=\textwidth]{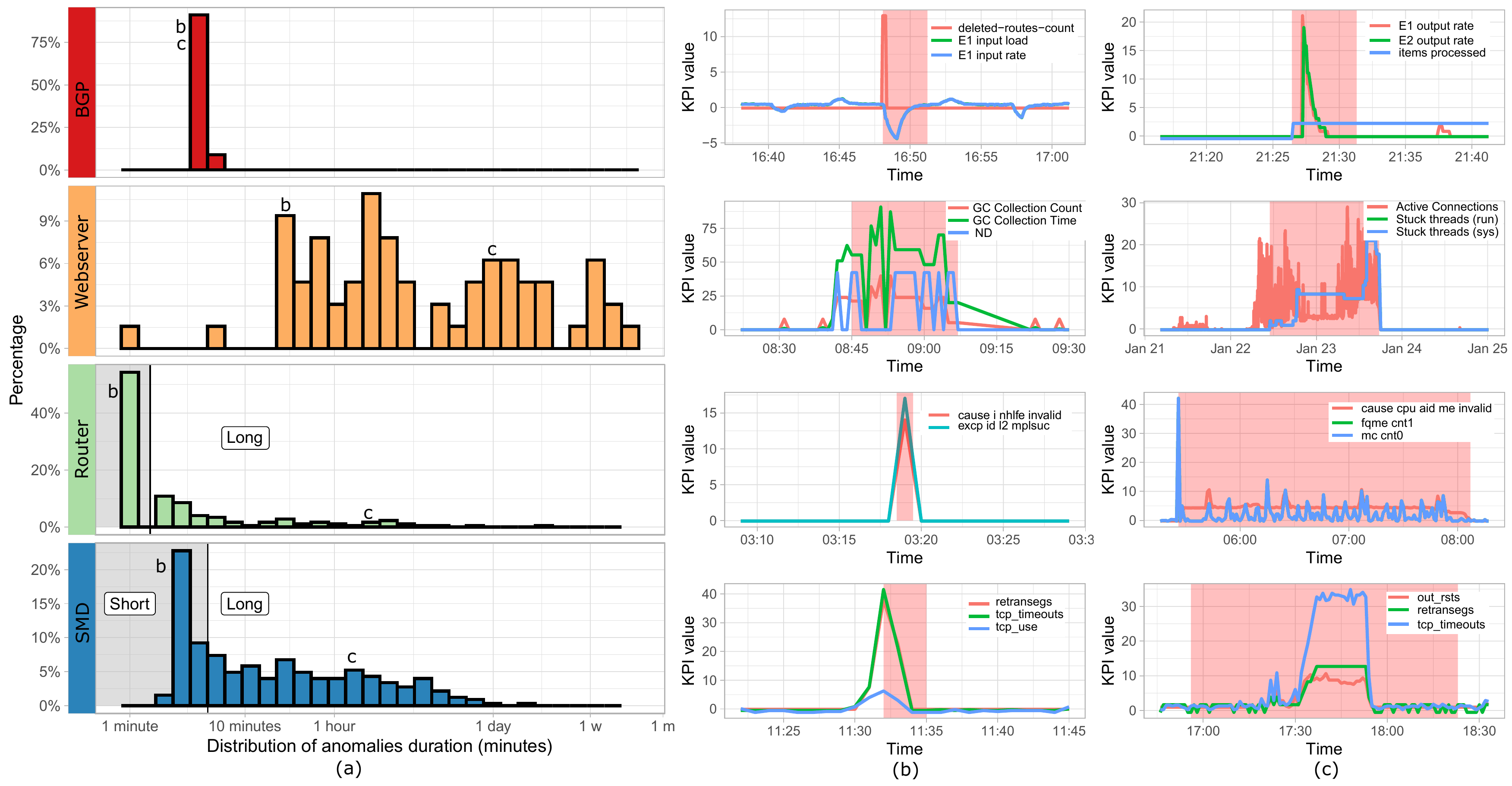}
    \caption{\emph{Temporal analysis}: (a) Distribution of the duration  anomalous events for all data sources. Anecdotal examples of anomalies with (b) short and (c)  long duration: standardized values of the top-3 KPIs  (lines)  and temporal GT (red ranges).}
    \label{fig:duration}
\end{figure*}


We start by analyzing the temporal properties of the anomalous GT. In this context, 
a KPI dataset with $F$ features and of duration $T$ samples can be represented as a $X \in \mathbb{R}^{T F}$  matrix corresponding to a multivariate numeric time series, and the corresponding ground truth is either (i) a temporal $GT \in \{0,1\}^{T}$   binary vector with $T$ values  indicating  whether the $t$-th time slot is anomalous, or (ii) a spatio-temporal  $GT \in  \{0,1\}^{T F}$  matrix describing  which  features are anomalous at any  sample. 
We start by characterizing the duration and interarrival of anomalous events in the GT.

\subsection{Anomalous events duration}
\firstfakepar{Goal} From a methodological standpoint, it is important to differentiate
\emph{outlier detection (OD)} defined as ``an observation
(or subset of observations) which appears to be inconsistent
with the remainder of that set of data''~\cite{barnett1979outliers}
from \emph{changepoint detection (CPD)}, whose focus is on ``detecting various
changes in the statistical properties of time series \ldots~such as
mean, variance, or spectral density''~\cite{gharghabi2017matrix}.  
Algorithms for OD and CPD  are specifically designed to capture these complementary types of
discrepancies and are hardly interchangeable: as such, it is important to observe the prevalence of events that are intrinsically better captured by either of OD or  CPD techniques. 

\fakepar{Method and insights} 
To simplify, we observe that CPD requires to observe change in statistical properties, which in turn require the observation to span multiple samples, and are thus only suitable for relatively \emph{long events}. Conversely, OD techniques are amenable to also detect \emph{short events}, such as spikes. To examine this question, Fig.~\ref{fig:duration}-(a) presents a histogram of the duration of the different anomalous events across data sources.

 The first clear observation is that each source has a completely different profile:  BGP  has a predefined anomalous event duration, and as such the delta-shaped distribution is expected and has limited interest. Webserver has an overall very long duration (about 9 months) and with few exceptions experts label events that last no less than 1 hour. Interestingly, no noticeable ``mode'' capturing a significant fraction of the event duration appear, so that events lasting 1 day or 1 week (i.e., more than 10K consecutive anomalous slots) are equally likely. Router and SMD instead exhibit a clear mode for short events: particularly, over 40\% (20\%) of events in Router (SMD) have a duration shorter than 1\,min (3\,min).
 At the same time, the rest of the events have a duration that spreads to really long windows of time (above 1\,day).
 
 Summarizing our observations, we first gather that  experts from the same technological domain may label anomalies in rather different manner: this may be due to an implicit bias to label short vs long duration, which are not fully elucidated by the terse representation of a binary ground truth.   Second, we see that a mixture of short outlier and long changepoint events are present across datasets: detecting both may require the simultaneous use of complementary OD and CPD techniques, which is rarely addressed in scientific research, where OD and CPD are separately studied.  
 
\fakepar{Additional GT observations}
Fig.~\ref{fig:duration} additionally portrays examples of anomalous events having (b) short  and (c) long  duration.
The picture  depicts the GT (red ranges) and the temporal evolution of the top-3 KPIs (according to their variation; in case of Router and SMD,  selection is limited to the top-3 KPI present in the anomalous GT). 

For BGP, the examples highlight issues that can arise with fixed ground truth: e.g., a fraction of the labeled window contains points that are similar or identical to periods without anomalies, which can yield false negatives in  evaluation.
This said, manual ground truth shows similar discrepancies. For instance,  in Webserver the GT appears to lag behind statistically anomalous behavior of the top-3 KPIs (for both short and long examples) while GT in the long event of SMD appears to fully include the event. Finally, it is interesting to contrast labeling of short anomalous events in Router (where a single sample is often labeled) vs SMD (where 3-samples, some of which exhibit normal behavior, surrounding the spike are often labeled).

Abstracting from these anecdotal observations, we gather that the time window labeled as anomalous does not appear to be  completely aligned with the anomalous looking data. This can have consequences on the quality of supervised models, but can also affect unsupervised evaluation as slot-based precision/recall metrics can be biased by the GT. To cope with this issue, a recent trend is the usage of evaluation metrics for temporal AD that allow for slight discrepancies (e.g., a flexible margin of error) between the ground truth and detections~\cite{xu2018unsupervised, hwang2019time}. At the same time, it appears that the GT labels in some data sources already ``hardcodes'' such a margin of error, for which the usage the aforementioned metrics would be discouraged. Thus, this calls for more systematic exposure of the processes of GT definition and universally adopted standard for AD evaluation.

\subsection{Anomalous event interarrival}

\firstfakepar{Goal} A complementary temporal property of anomalous events is the \emph{duration of the normal period between events}, i.e., otherwise stated, the interarrival time of anomalous events. Due to the diversity of the data sources, we are not interested here in a precise characterization of GT labels as a stochastic  process. 
Rather, our interest is in gathering implications for AD tasks.
For multivariate series with several anomalies, the temporal distance between them has implications regarding detection: e.g., if two independent anomalies are close in time, there is the risk of jointly detecting them as a single event, potentially misunderstanding both events.  

\fakepar{Method and insights}
More formally, given a  ground truth  vector GT for a specific dataset $d\in\mathcal{D}$ in the data source $\mathcal{D}$, we extract from the ground truth all anomalous events  as tuples $(d, s_i, e_i)$,  where $s_i$ and $e_i$ represent the start and end-time of the $i$-th event, such that  $(GT_t=1, \forall t \in [s_i,e_i]) \wedge (GT_{s_i-1} = GT_{e_i+1} = 0)$.   
Considering ordered tuples $s_i>e_{j},  \forall i>j$, we then  define the interarrival time as the difference between consecutive $s_i - e_{i-1}$ events in a given dataset $d\in\mathcal{D}$.

\begin{figure}
    \centering
    \includegraphics[width = \columnwidth]{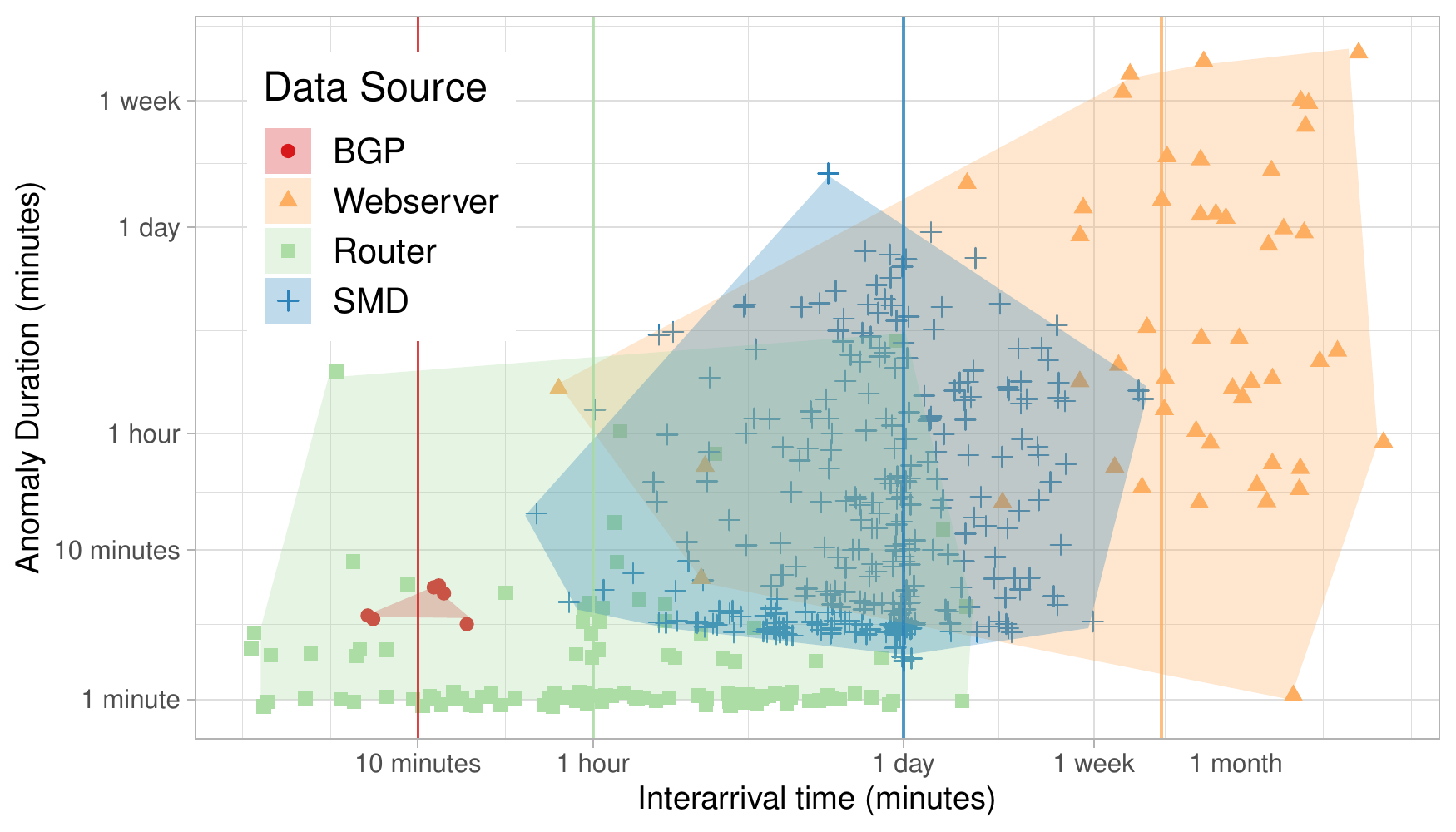}
    \caption{\emph{Temporal analysis}: Scatter plot of interarrival time (x-axis) versus anomaly duration (y-axis), with median lines.}
    \label{fig:inter_vs_dur}
\end{figure}

Fig.~\ref{fig:inter_vs_dur} depicts the scatter plot of anomaly interarrival time  versus anomaly duration (reporting for the sake of visibility, a colored convex hull comprising all points in a given data source), annotating the median interarrival with a vertical line.   As for the anomaly duration, we observe that also the anomaly inter-arrival process is highly variable across datasets: in Webserver anomalies are rare events (median interarrival of two weeks), while the labeling timescales of  SMD (median interarrival 1 day)  and  Router (median interarrival 1 hour) sources are significantly shorter, and BGP anomaly injections appears to be periodic (around 10 minutes).  In particular,  the distribution of points and the extent of the hulls in Fig.~\ref{fig:inter_vs_dur}  show that for most sources, there is no systematic bias in the anomalous arrival pattern (e.g, long anomalies followed by long interarrival, or  short anomalies one after the other).
At the same time, it appears that for datasets with synthetic GT (such as BGP), trivial rules learned by ML on data (such as a detector  learning to raise 5-minutes long alarms every 10-minutes) could appear to be ``good for the wrong reasons''~\cite{lapuschkin2019unmasking} --- while such oddities are unlikely in the more complex situation presented by human labeling.


   

\section{Spatial Properties}\label{sec:result_space}
%
    
 
We next turn our attention to the analysis of the spatial component of the ground truth labels, which is possible for the SMD and Router sources.  We first quantify the spatial footprint of anomalous events. We next use clustering techniques to automatically infer an anomalous type classification and study their relative occurrence.

\subsection{Spatial footprint of anomalous events}\label{sec:spatial_footprint}

\firstfakepar{Goal} The fraction of KPIs related to an anomaly influences the difficulty of detecting it. It has to be noted that multivariate AD methods~\cite{truong2020selective,if08icdm,hariri2019extended,pevny2016loda,manzoor2018xstream,kwon2019survey} are  designed to produce a single anomalous score for the whole event.  As such, the identification of the spatial components that most contribute to making the timeslot anomalous are computed by entirely other means (e.g., via reconstruction error~\cite{sota_ad_dl1}, by evaluating feature deviation~\cite{sota_other_adele}, feature importance~\cite{sota_lookout} or feature predictive power~\cite{PROTEUS}).
Additionally, several state of the art methods~\cite{pevny2016loda,manzoor2018xstream} are based on sparse random feature projections: thus, a low percentage of anomalous KPIs, paired with a high number of KPIs, can make correct spatial detection a hard task. 

\fakepar{Method and insights}  
As previously done for the temporal characteristics, we simplify the task and start by 
contrasting intrinsically \emph{univariate} anomalies (i.e., for a given $t$, only a single KPI $j$ is anomalous $\exists ! j: GT_{t,j}=1 $) vs anomalies with a  \emph{multivariate} footprint (where for a given $t$  we quantify the footprint as the number of anomalous KPIs in the GT    $| j: GT_{t,j}=1 |$, possibly normalized over the number $F$ of KPIs  in the dataset).

\begin{figure}[t]
 \centering
 \includegraphics[width=.99\linewidth]{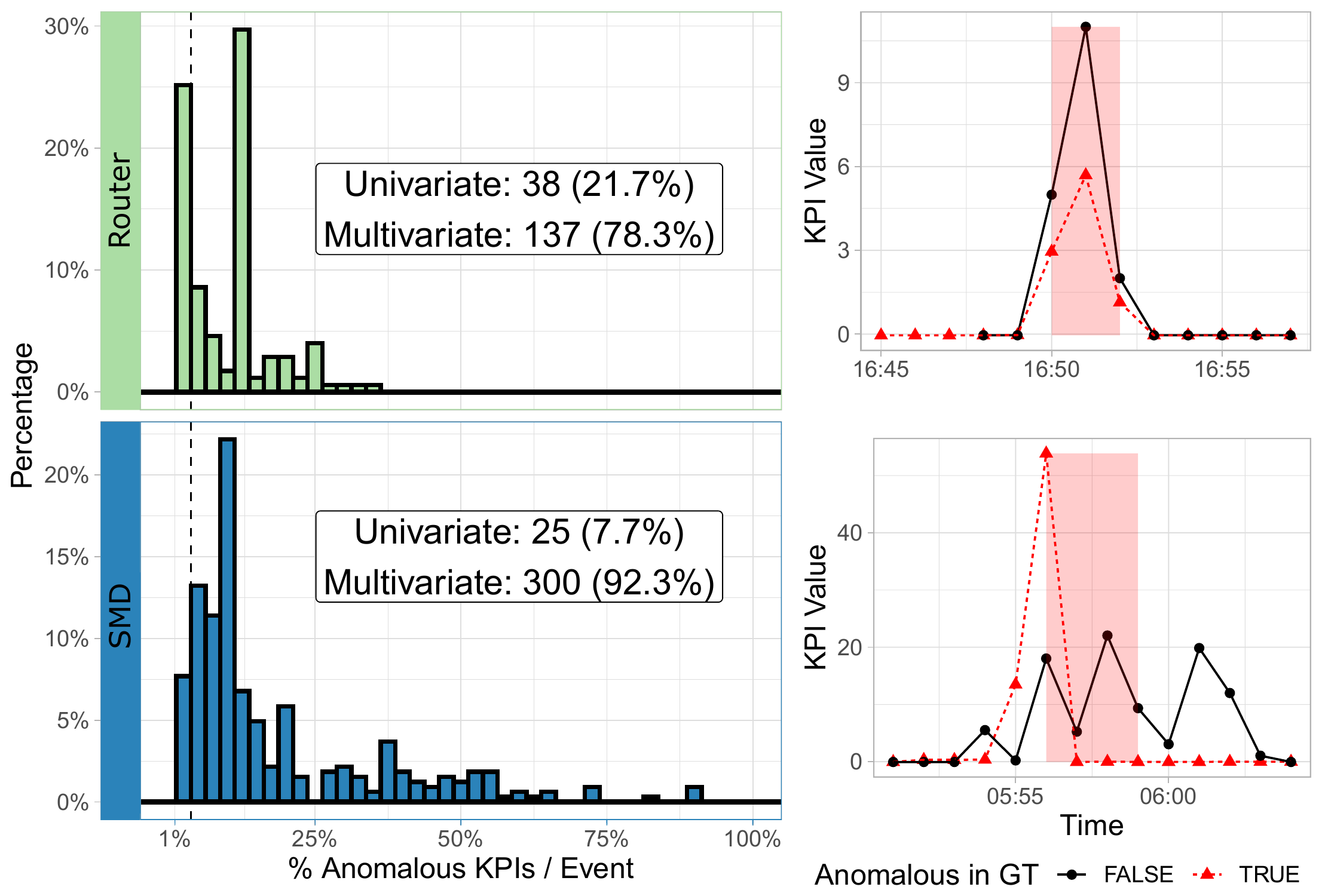}
\caption{\emph{Spatial analysis}: (a) Distribution of the percentage of anomalous KPIs per-event (the dashed line separates univariate from multivariate anomalies) and (b) Examples of univariate anomalies (KPI $\in$ GT in dashed red, additional KPI  $\notin$ GT   exhibiting anomalous behavior  in solid black).}
\label{fig:kpis}
\end{figure}

Fig.~\ref{fig:kpis}-(a) reports a histogram of the percentage  of involved KPIs  per anomalous event, which shows the eminently  multivariate nature of human labeled anomalies: only 7\% (21\%) of SMD (Router) events are univariate. Additionally, while the  median number of anomalous KPIs per event is 4, the events with the largest footprint comprise up to few tens of anomalous KPIs. At the same time, the tail in Fig.~\ref{fig:kpis}-(a) shows that as a small fraction of KPIs is labeled, events do not tend to disrupt the whole system: rather, a small fraction of KPIs is possibly enough to characterize the event type -- which deserves further attention.

\fakepar{Additional GT insights} Anecdotal examples of  univariate anomalies are reported in Fig.~\ref{fig:kpis}-(b), where  KPIs labeled as anomalous are plotted in a red dashed line, while unlabeled KPIs are presented in solid black line: in both cases, we can observe that KPIs that exhibit a statistically anomalous behavior are not labeled as such by the expert that examined them.
On the one hand, this suggests that the fraction of purely univariate (from a statistical viewpoint) anomalies is overestimated in the ground truth. On the other hand, experts may knowingly discard KPIs that  are not  relevant from a business or domain viewpoint: it appears that some amount of expert knowledge, which is complementary to KPI data and thus not readily available to the algorithm, needs to be considered to reconcile algorithmic and expert viewpoints.

\begin{figure}
    \centering
    \includegraphics[width = \columnwidth]{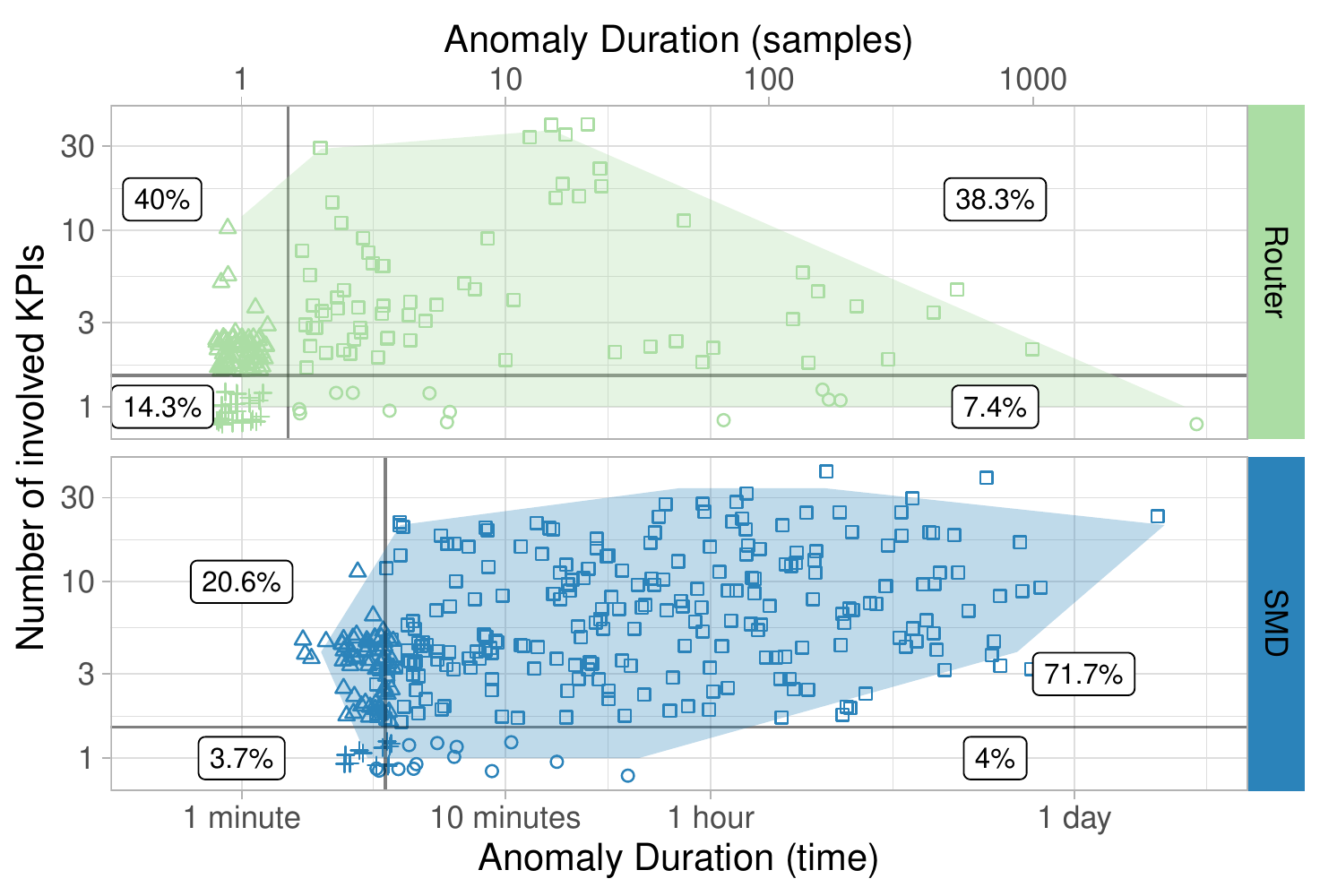}
    \caption{\emph{Spatio-temporal analysis}: Jittered scatter plot of anomaly duration vs spatial footprint. Lines separate the quadrants defining temporal (short vs long duration) and spatial  (univariate vs multivariate) characteristics of the GT.}
    \label{fig:dur_vs_kpis}
\end{figure}

\fakepar{Spatio-temporal insights}
Finally, we provide a joint spatio-temporal view  and examine the duration of anomalies vs their spatial footprint on Fig.~\ref{fig:dur_vs_kpis}. Duration and number of involved KPIs appears to be unrelated (Pearson correlation tops to 0.22 for SMD), it is interesting to note that all cross-product of (short, long) and (uni, multi)-variate anomalies are present in the sources.   Additionally, both sources present a non marginal fraction of anomalies with low numbers of anomalous KPIs but extremely long durations, which are particularly hard to detect. 

\begin{figure*}
    \centering
    \includegraphics[width = 2\columnwidth]{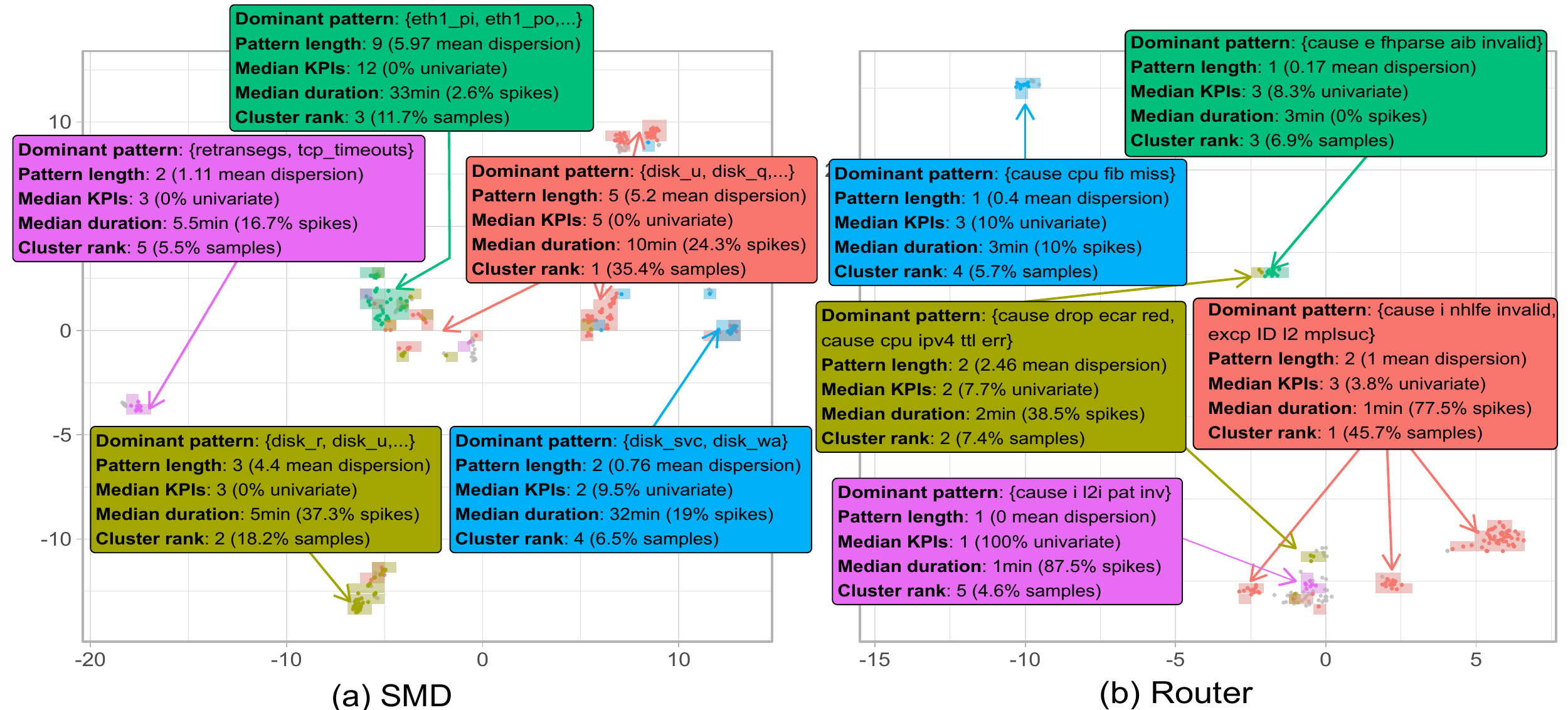}
    \caption{\emph{Spatial analysis}: Representation of anomalous event clusters, projected in a
     2d space, for (a) SMD and (b) Router. Annotations summarize  characteristics of  the top-5 clusters obtained by Proximus. }
    \label{fig:proximus}
\end{figure*}

\subsection{Popularity of anomalous event types}\label{sec:results_anomalydb}
\firstfakepar{Goal}
The preliminary characterization of the spatial  footprint of an anomaly  suggests that human experts characterize anomalies using a small fraction of KPIs, which can be used to group together anomalies presenting similar sets of anomalous KPIs. The previous characterization, though, presented an amount of noise, bias and even possible errors in the ground truth labels, which any  grouping process should be aware of. 
Our   goals in this section are thus to assess whether such KPIs groups emerge from  the GT data via clustering and to additionally inspect properties of resulting clusters. 

\fakepar{Method}
For grouping together anomalies with similar set of KPIs, we resort to clustering in a \emph{binary}  (as each KPI is either present of absent), \emph{large} (38 dimensions for SMD and over 100 for Router)  and \emph{sparse} (as the median number of KPIs per anomaly is 4) space.
More formally, for each dataset in a source $\mathcal{D}$, given an anomalous event $i \in [1,N]$ (with $N$ the overall number of anomalous events), starting at date $s_i$ and ending at $e_i$, we consider the multivariate set of  anomalous features $k_i = \{ j : \exists t \in [s_i, e_i], GT_{t,j} = 1 \}$. Let further denote with $F$ the number of unique features across all events, as gathered by one-hot encoding of features. We then construct the matrix $K \in \{0,1\}^{NF}$ whose elements $K_{i,j}=1$ when anomaly $i$ contains feature $j$, and $K_{i,j}=0$ otherwise. 
Due to aforementioned noise, labeling imprecision etc., we refrain from just clustering exactly matching footprints (i.e., equal rows in the matrix) and rather seek for approximate matches in this 
binary, sparse, high dimensional space. To do so, we  employ Proximus~\cite{proximus},  a binary clustering algorithm, that approximates the original matrix with sets of dense submatrices. Furthermore, a noticeable feature of Proximus is that clusters are self-described by a \emph{dominant pattern}, i.e., a small set of anomalous features that uniquely identifies and represents the cluster.
Proximus  depends on two parameters, namely: (i) the \emph{maximum radius  $R^+$}, i.e., expressed as the maximum number of bits a member in a cluster may deviate from its dominant pattern and (ii) the \emph{minimum size  $S^-$}, i.e., the minimum number of samples in a cluster.

Clearly, while the algorithm attempts to create clusters that satisfy the imposed conditions, in some cases conditions are violated: in this case, Proximus creates a cluster with an ``empty'' dominant pattern, which is used as a rag bag~\cite{amigo2009comparison},  in which unassigned samples are grouped together. To fix the hyperparameters, we performed for each data source a grid experiment where we tested all possible combinations of $R^+ \in [0, F]$ and $S^- \in [1,N]$, and selected the resulting clustering that minimized, in this order: (i) the size of  the rag bag cluster, i.e., the total number of samples without dominant pattern, (ii) the total approximation error, i.e., the sum of distances from each sample to its dominant pattern over all samples, (iii)  $R^+$ and $S^-$. Notice this order  ensures that (i) most of the samples are clustered, that (ii) clustering process provides consistent results, as samples are close to their dominant pattern and (iii) there are as many clusters as possible, to provide fine-grained anomaly footprint. 
 
\fakepar{Insights on cluster composition} 
We applied the presented methodology to the ground truth of both the Router and SMD sources. A bi-dimensional visualization of the top-5 resulting clusters is reported in Fig.~\ref{fig:proximus}.

We note that the top-5 clusters  represent  70\% (77\%) of the Router (SMD) events:  clusters correspond to different types\footnote{Recall that clustering happens in a much larger space, so clusters are only apparently close due to 2d projection.}  of anomalies, which are automatically-labeled via Proximus' dominant pattern (of which only up the first two components are reported in the figure due to space limits). For each cluster, annotations further characterize: the dominant pattern length and mean dispersion (i.e., the average value of the Manhattan distance between members of each cluster and its pattern); the median number of KPIs (and percentage of univariate events); the  median duration (and percentage of spikes); the cluster rank and size (percentage of overall anomalies).
We remark that as the mean dispersion of samples in the clusters is small, it follows that clusters represent a consistent set of highly similar KPIs, testifying to the soundness of the approach.

Due to space limitations, we avoid reporting a full description of clusters, that would be interesting for domain experts but does not add value from a methodological viewpoint.  
We instead provide a mathematical model to further testify  that clusters provided by our method are not artifacts.
In particular, we do so by showing that the expert labelling exhibits patterns that would not appear for randomly generated data. Assuming two events generated by taking randomly $A$ anomalous KPIs among $F$ KPIs, the number of shared KPIs $l$ verifies:
\begin{align}
\mathbb{P}(l) = \binom{A}{l} \binom{F-A}{A-l} \bigg/ \binom{F}{A},
\end{align}
giving an expected number $\mathbb{E}[l]$  of shared KPIs:
\begin{align}
\mathbb{E}[l] = \sum_{l=0}^{F} l P(l) = \sum_{l=0}^{F} l \binom{A}{l} \binom{F-A}{A-l} \bigg/ \binom{F}{A} = A^2/F.
\end{align}
Numerically, taking $A=4$ (median number of anomalous KPIs for both Router and SMD) and either $F=87$ (Router) or $F=38$ (SMD), \emph{we expect less than one KPI in common} in case of random labeling. Therefore, observing large groups of spatial clusters as those reported by Proximus,  indicates that the labeling performed by the expert is exhibiting a semantically relevant structure.


\begin{figure}[t]
    \centering
    \includegraphics[width = 0.9\linewidth]{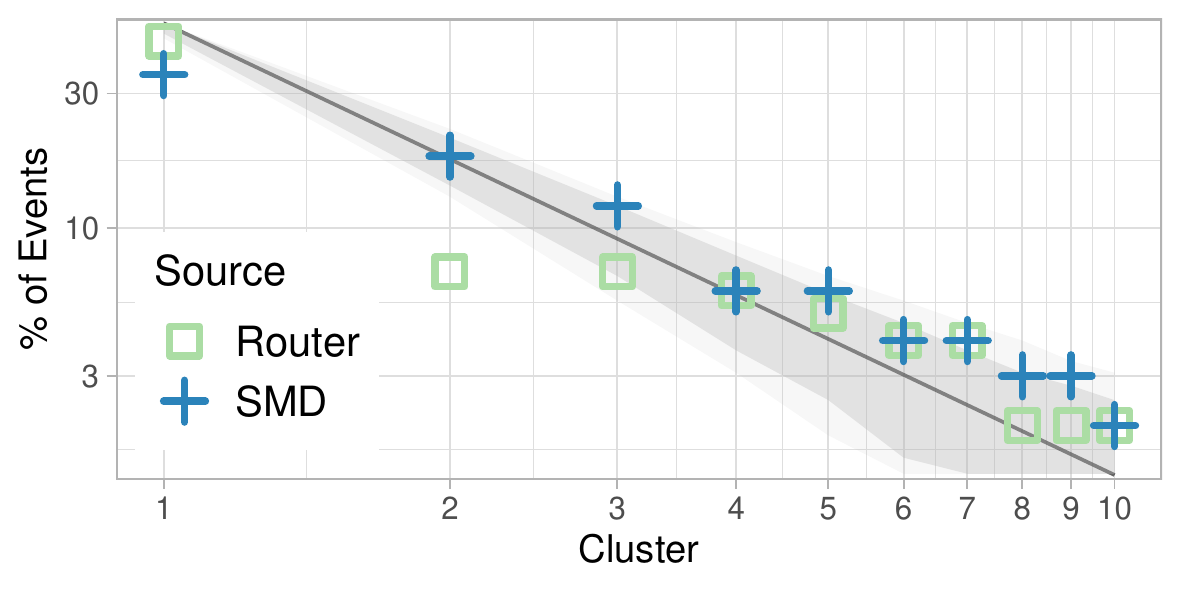}
    \caption{\emph{Cluster popularity}: Relative popularity of different anomaly types, as identified by their cluster ID. Heavy tailed popularity distribution is reported for visual comparison only (Zipf with catalog $N$ = 10, shape $\alpha$ = 1.6; shades  representing the confidence interval at $95$ and $99\%$ for a set of $325$ events).}
    \label{fig:zipf}
\end{figure}

\fakepar{Additional insights on cluster popularity} 
As early observed, cluster size has a skewed distribution: 
while the overall number of clusters is 30 (18) for Router (SMD), roughly 3/4 of amomalous samples falls in just the top-5 clusters, whereas top-10 cluster comprise 85\% (93\%) of all events in the Router (SMD) dataset.
 In other words, despite anomalies are rare events in the time dimension, some anomaly types are even rarer than others in the spatial dimension. We depict the popularity skew by showing the percentage of events in each cluster, for the top-10 clusters ranked by decreasing size in the log-log plot of Fig.~\ref{fig:zipf}. The picture shows that relative popularity of anomalies follows a highly skewed  popularity distribution\footnote{Note that, due to the small catalog and sample size, it would not be statistically relevant to attempt to fit a precise distribution: the Zipf law and simulation envelope for finite realizations of the same length is merely added for visual reference.}, which has important consequences for data collection, labeling and operation.

First, we stress that popularity results are remarkably similar over two independent data sources, collected in distinct environments with no overlap in the  KPI signals,  heterogeneous spatio-temporal characteristics and rather different labeling strategies. This reinforces the soundness of our approach and the generality of the findings concerning the existence of a remarkable popularity skew across anomaly types.

Second, this means that the existence of such skew can be leveraged to gather
a finer grained understanding of popular anomaly types. For instance, we previously noticed that not all KPIs that are statistically anomalous are labeled by experts: should that happen systematically for some specific KPI in a cluster, this could help automatically inferring the low business relevance for that KPIs and anomaly type, i.e., semi-supervised expert knowledge can be encoded at cluster level. Similarly, it is possible that, for some clusters, anomalies may share the same cause and exhibit similar effects: this could simplify the  root cause analysis process, which should be done once per cluster (as opposed to once per event).

Third, while in this paper we apply Proximus in offline batch mode, it can be argued that  clustering can be applied online in incremental fashion, using any of~\cite{sota_ad_dl1,sota_other_adele,sota_lookout,PROTEUS} and  spatially projecting  unlabeled anomalies in the same space,  labeling only new samples that do not fit existing clusters. Indeed, after few samples of the same cluster are labeled by the expert, there is no need to further systematically label all events from the same cluster -- which, as anomalies are skewed, would allow to significantly reduce the labeling effort. Moreover, due to popularity skew, we can expect to quickly reach a good characterization of few clusters representing the majority of events.


\subsection{Gain of cluster-based active-labeling}
\firstfakepar{Goal} We observe that spatial skew can yield to a reduction of labeling effort, that we aim at quantifying. As opposite as to present a full-blown active-labeling system, we aim at upper-bounding the gains that a stratified selection of samples based on clustering could bring.  We instead leave the study of online algorithms to approach such gains for future work.

\begin{figure}[t]
 \centering
 \includegraphics[width=.9\linewidth]{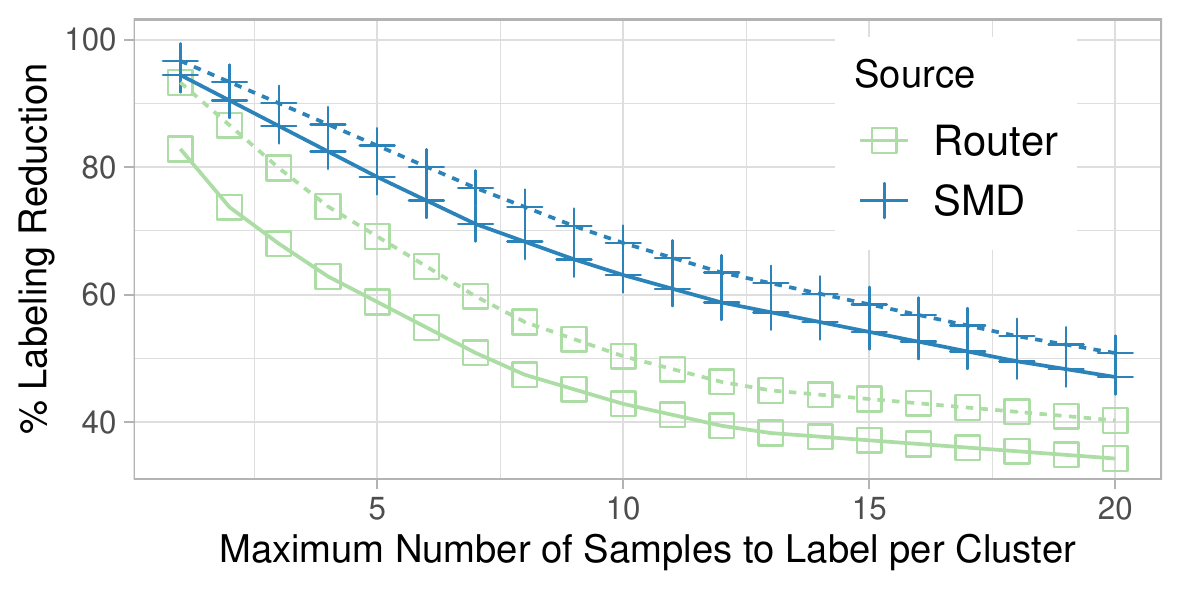}
\caption{\emph{Gain of clustering-driven active labeling}: reduction of labeling effort  as a function of the number of samples to label per cluster (dashed lines correspond to labeling only samples in the top-10 clusters, whereas solid lines correspond to all clusters).}
\label{fig:active_labeling}
\end{figure}

\fakepar{Method and insights} 
We leverage the exhaustive set of labels from $N$ anomalous events, which required human expertise, and assess how many of these $N$ labels are \emph{redundant} and could have been avoided, as they bring little additional information. As per the observed popularity skew of the spatial anomaly distribution, the bulk of the events  can be categorized into a small set of anomalous types  $C$ that can be gathered by clustering.

Observe that due to the skew, many labels of the popular anomalous types/clusters are redundant: e.g., with Zipf skew $\alpha=1$, samples of the most popular anomaly type/cluster are $k$ times more frequent labeled than for the $k$-th type/cluster -- and the redundance grows with $\alpha$, that as we early shown appears to exceed 1 in our datasets. Note that this Zipf comparison is heuristic and, in our data, we expect the smaller clusters to be of size 1 as N increases.

By using clustering, the labeling effort can be significantly reduced by capping the amount of samples per cluster to a maximum $M$, as opposite to as systematically labeling all anomalous events.
As we aim at upper-bounding the active labeling gain, we consider an ideal algorithm creating spatial footprints that exhibit a perfect agreement with expert labeling. Yet, we point out that 
our previous work~\cite{itc32} has shown effective unsupervised (and semi-supervised) algorithms for 
automatically gathering spatial anomalous footprint, that exhibit high levels of agreement with expert labeling, so that we expect that it would be possible to design practical strategies to approach such upper-bound with even less human intervention.

Fig.~\ref{fig:active_labeling} quantitatively upper-bounds the expected reduction by considering two active-labeling strategies, where either (i) the expert labels only $M$ samples for the top-10 clusters or (ii) the expert labels $M$ samples for all clusters, for varying $M\in[1,20]$.
 The picture clearly shows that, depending on the number of labeled samples, the labeling effort can decrease by \emph{up to an order of magnitude} (e.g.,  when $M\approx 5$ per cluster are labeled in the  SMD data source). Moreover, the expected labeling reduction is any case at least a factor of 2  (i.e., for any strategy and data source). Even with this rather conservative estimate, the labeling effort  would already be \emph{halved}:  in turn, this would translate into either a cost reduction (e.g., less time/experts), or an expected increase in the labeling quality (by focusing the expert attention on useful labels).

\section{Conclusions}\label{sec:discuss} 

In this paper, we carry on a systematic analysis of Ground Truth (GT) available for (i) multi-variate (ii) temporal data for network troubleshooting using four data sources.
While the importance of GT for temporal data is discussed in \cite{wu2021current}, two of the four considered data sources enabled us to analyze labeling from a (iii) spatial viewpoint, which to the best of our knowledge we are the first to analyze.

In particular, we show that (i) anomalies exhibit distinct spatial patterns that are not due to chance: we propose to leverage a clustering technique that can \emph{assist human labeling}, by directly providing  \emph{dominant patterns} of similar events.  
We further show  that (ii) albeit anomalies may be rare over time, their \emph{spatial pattern exhibit a significant popularity skew}:  
interestingly, despite the environment, the collected KPIs, the labeling process and the temporal patterns exhibiting significant differences across the two sources, our analysis gathered consistent remarks at both qualitative and quantitative levels.
Finally, we show that (iii) by \emph{exploiting clustering of the spatial anomaly footprint as a means to perform stratified sampling in active-labeling settings}, one can achieve significant reduction of the expected labeling effort  In particular, according to the data sources at our disposal,  we observe that reduction is of at least 2$\times$ (and up to 10$\times$), which can be beneficial to, e.g.,  reduce GT labeling cost and improving GT quality overall.

As future work, we plan to address the design and implementation of practical online strategies, with the goal of building larger and high-quality domain-specific anomaly databases,  with a sustainable labeling effort.

\bibliographystyle{IEEEtran}

\bibliography{refs}

\end{document}